\documentclass[sigconf,screen]{acmart}

\usepackage{graphicx}
\usepackage{textcomp}
\usepackage{tcolorbox}
\usepackage[]{xcolor} 
\usepackage{cleveref}
\usepackage{multirow}
\usepackage{multicol}
\usepackage{booktabs}
\usepackage{pifont}
\usepackage{bbding}
\usepackage{fontawesome}
\usepackage{xinttools}
\usepackage{arydshln}
\usepackage{makecell}
\usepackage{tabularx}
\usepackage{cellspace}
\usepackage{booktabs}
\usepackage{array}
\usepackage{caption}
\usepackage{rotating}
\usepackage{amsmath}
\usepackage[utf8]{inputenc}
\usepackage{balance}
\usepackage{adjustbox} 
\usepackage[commandnameprefix=always]{changes}

\AtBeginDocument{%
  }

\definecolor{log_example_blue}{RGB}{254,192,0}

\newcommand{\answerbox}[1]{%
    \par
    \noindent%
    \setlength{\fboxrule}{1pt}%
    \fcolorbox{gray!75!black}{gray!15}{%
        \parbox{\dimexpr\linewidth-2\fboxsep-2\fboxrule}{%
            \normalsize #1%
        }%
    }%
    \par%
}

\newcommand{\otherbox}[1]{%
    \par
    \noindent%
    \setlength{\fboxrule}{1pt}%
    \fcolorbox{orange!75!black}{orange!5!white}{%
        \parbox{\dimexpr\linewidth-2\fboxsep-2\fboxrule}{%
            \normalsize #1%
        }%
    }%
    \par%
}

\copyrightyear{2026}
\acmYear{2026}
\setcopyright{cc}
\setcctype{by}
\acmConference[ICSE '26]{2026 IEEE/ACM 48th International Conference on Software Engineering}{April 12--18, 2026}{Rio de Janeiro, Brazil}
\acmBooktitle{2026 IEEE/ACM 48th International Conference on Software Engineering (ICSE '26), April 12--18, 2026, Rio de Janeiro, Brazil}
\acmPrice{}
\acmDOI{10.1145/3744916.3787801}
\acmISBN{979-8-4007-2025-3/2026/04}

\begin{document}

\title{LogFold: Compressing Logs with Structured Tokens and Hybrid Encoding}


\author{Shiwen Shan$^{\P}$, Yintong Huo$^{\ddag}$, Hongzhan Zhong$^{\P}$, Zhining Wang$^{\P}$, Yuxin Su$^{\P*}$, Zibin Zheng$^{\P}$}
\affiliation{%
  \institution{$^{\P}$School of Software Engineering, Sun Yat-sen University, $^{\ddag}$Singapore Management University}
  \country{}
}
\email{{shanshw, zhonghzh8, wangzhn23}@mail2.sysu.edu.cn}
\email{ythuo@smu.edu.sg, {suyx35,  zhzibin}@mail.sysu.edu.cn}

\thanks{* Yuxin Su is corresponding author.}

\renewcommand{\shortauthors}{Shiwen Shan, et al.}

\begin{abstract}
Logs are essential for diagnosing failures and conducting retrospective studies, leading many software organizations to retain log messages for a long time.
Nevertheless, the volume of generated log data grows rapidly as software systems grow, necessitating an effective compression method.
Apart from general-purpose compressors (e.g., Gzip, Bzip2), many recent studies developed log-specific compression algorithms, but they offer suboptimal performance because of (1) overlooking redundancies within certain complex tokens, and (2) lacking a fine-grained encoding strategy for diverse token types. 

This work uncovers a new redundancy pattern in structured tokens and proposes a new type-aware encoding strategy to improve log compression. 
Building on this insight, we introduce \texttt{LogFold}, a novel log compression method consisting of four components: a token analyzer to classifies tokens as structured, unstructured, or static types; a processor that mines recurring patterns within structured tokens based on their delimiter skeletons; a hybrid encoder that tailors data representation according to token types; and a packer that compresses the output into an archive file.
Extensive experiments on 16 public log datasets demonstrate that \texttt{LogFold} surpasses state-of-the-art baselines, achieving average compression ratio improvements by 11.11\%, with a compression speed of 9.842 MB/s. Ablation studies further indicate the importance of each component. We also conduct sensitivity analyses to verify \texttt{LogFold}’s robustness and stability across various internal settings.

\end{abstract}

\ccsdesc[300]{Software and its engineering~Software maintenance tools}


\keywords{data compression, log compression, log management, log analysis}
\maketitle

\section{Introduction}
System logs are an indispensable resource in software engineering. They provide valuable run-time information about software behavior for system maintenance~\cite{du2017deeplog,yang2021semi,zhang2019robust}, ~\cite{shan2024face,huang2024demystifying,jiang2025l4}, performance modeling~\cite{agrawal2018log,yao2018log4perf}, and security auditing~\cite{ahmad2022hardlog,sekar2024eaudit}. 
Additionally, the detailed information in logs necessitates their storage for extended periods to support post-mortem analysis~\cite{ding2012ErrLog, jia2018smartlog,he2022empirical,miranskyy2016operational}. 
Industry practices have regulations on mandating long-term log storage~\cite{wei2021feasibility,wang2024muslope}.
For example, some cloud providers must store logs for 180 days~\cite{wei2021feasibility}, and others are required to keep them for years~\cite{rodrigues2021clp}. 
Legal compliance, such as data privacy laws, also mandates the prolonged storage of user logs~\cite{yao2020study}.

However, as software systems scale, they generate an enormous volume of logs, imposing huge pressure on storage systems in terms of cost and capacity~\cite{spillner2020comparison}. For instance, large-scale services like eBay were reported to produce over a petabyte of log data daily in 2018~\cite{rodrigues2021clp,wang2024muslope}.
Consequently, to mitigate the challenge, effective log compression has become an indispensable technique. 

The core idea of log compression is to \textbf{identify} recurring patterns~(e.g., repeated templates, common tokens) within the log data and 
\textbf{encode} these patterns and the remaining data into a compact form to eliminate redundancy.
Here, \textit{encoding} refers to the transformation of original data into a more space-efficient representation.

To compress logs, a straightforward solution is to employ general-purpose compressors such as Gzip~\cite{gzip} and Bzip2~\cite{bzip2}.
However, their compression ratios remain suboptimal because they fail to recognize the various recurring patterns~(e.g., highly repeated log templates) of log~\cite{yao2020study, wei2021feasibility, rodrigues2021clp}.
To further improve log compression, a variety of specialized log compressors have been proposed~\cite{liu2019logzip, li2024logshrink, yu2024unlocking,feng2016mlc,yao2021improving}. 
The parsing-based log compressors typically target reducing redundancy in log templates (i.e., log events). To extract recurring templates, LogZip~\cite{liu2019logzip} leverages an iterative-clustering method, while LogReducer~\cite{wei2021feasibility} and LogShrink~\cite{li2024logshrink} utilize sample-based log parsing techniques. 
Then, all three methods employ dictionary encoding to replace templates and their parameters with compact indexes. Additionally, LogReducer enhances compression by applying enhanced numeric encoding strategies~(e.g., elastic encoding and delta encoding) to number sequences.
In contrast, Denum~\cite{yu2024unlocking} compresses numeric tokens by grouping them based on length and initial digit, followed by dictionary and optimized numeric encoding.
While effective, these approaches face two key limitations:
(1) overlooking the recurring patterns within \textit{structured tokens}.
and (2) lacking a comprehensive, type-aware encoding strategy for handling \textit{token diversity}.

The first limitation is the oversight of recurring patterns within structured tokens. Log entries often include tokens composed of delimiters and sub-tokens, such as \texttt{2015-07-09}.
Current methods typically do not tailor to these more complex structured tokens unless they match exactly. 
However, these tokens still contain recurring patterns inside, such as the format \texttt{<>-<>-<>}, and the sequence of sub-tokens like \texttt{["2015", "07”, “09”]}. Ignorance of such intra-token characteristics will hamper the effectiveness of compression.

The second limitation is the lack of a comprehensive, type-aware encoding strategy. Log entries contain diverse token types: pure numbers, pure strings, and, most commonly, mixed-type tokens~(e.g., a background service process's name and its ID, such as \texttt{ftpd[4305]}). However, existing works employ a coarse approach, which applies numeric encoding only to pure numbers while treating pure strings and complex mixed-type tokens alike with plain dictionary encoding. This is sub-optimal, as it ignores the distinct structural characteristics of mixed-type tokens, thereby sacrificing compact representations for encoding diverse tokens.

This paper stems from two insights to mitigate the above limitations. 
First, we recognize and decompose structured tokens based on their delimiter skeletons and sub-token sequences. The deconstruction uncovers novel intra-token patterns, which can be used to enhance compression.
Second, we emphasize a fine-grained hybrid encoding strategy that tailors effective encoding techniques to each specific data type.
This holistic view ensures that all tokens~(i.e., numeric, string, and mixed) are compressed effectively, leading to overall compression improvements.

We propose and implement \texttt{LogFold} to capitalize on these insights.
It is a delimiter skeleton-aware log compression tool consisting of four components.
Specifically, \texttt{LogFold} begins with a \textit{Token Analyzer} that treats each log entry as a whole.
Rather than breaking it apart, the analyzer preserves the original sequence and classifies tokens into three types: structured, unstructured, and static.
The identified structured tokens are then passed to the \textit{Structured Token Processor} designed to exploit the recurring patterns within structured tokens. 
It begins by deconstructing tokens and grouping them according to their shared delimiter skeleton, reducing redundancy in structural patterns.
Next, it applies pattern mining to sub-tokens within those groups, recursively partitioning them into smaller, more homogeneous subgroups. 
This two-step strategy, progressing from coarse-grained skeletons to fine-grained sub-tokens, enables significantly identifying recurring patterns, and thus enhances compression.
Then the resulting refined patterns and the simplified sub-token matrix, along with unstructured tokens and the static token sequences, will be passed to the \textit{Hybrid Encoder} component.
This component utilizes two optimized classic encoding strategies and a novel proposed mixed-type encoding strategy to encode the input data for higher compression gains. 
Finally, the encoded files, including dictionaries and binary data, are packed and compressed by a Packer using a general-purpose compressor to produce the final compressed archive.

Evaluations on 16 public log datasets demonstrate the capability of \texttt{LogFold} to outperform nine baseline compression tools by 11.11\% to 494.19\% in terms of average compression ratio, achieving the highest compression ratio on 12 out of 16 datasets, with a compression speed of 9.842 MB/s.
Further, the results of the ablation study show the individual component contributions to \texttt{LogFold}.
We also conduct a comprehensive sensitivity experiment to show the robustness and stability of \texttt{LogFold}.

To conclude, the main contributions are listed as follows:

$\bullet$ We identify two limitations in existing log compressors: neglecting structured token details and lacking a comprehensive encoding strategy.

$\bullet$ We propose \texttt{LogFold}\footnote{Publicly available in \url{https://github.com/shanshw/LogFold}}, a skeleton-aware log compressor with a hybrid encoding strategy that captures recurring patterns within structured tokens and encodes data based on specific token types.

$\bullet$
We demonstrate the superior performance of \texttt{LogFold} through extensive evaluations on 16 public datasets, achieving an average compression ratio improvement of 11.11\% over state-of-the-art methods while maintaining a competitive speed.

\section{Preliminaries}
\begin{figure}
    \centering
    \includegraphics[width=0.7\linewidth]{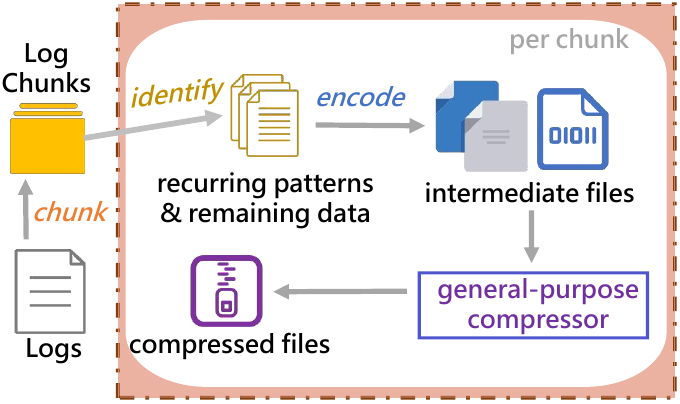}
    \vspace{-0.1in}
    \caption{The general log compressor paradigm.}
    \vspace{-0.2in}
    \label{fig:log-compression-paradigm}
\end{figure}
\subsection{Log Compression}
General-purpose compressors aim to eliminate generic recurring patterns across diverse data sources but often fail to exploit the distinct characteristics in logs~\cite{yao2020study, liu2019logzip, feng2016mlc}. Dictionary-based methods like Gzip~\cite{gzip} and LZMA~\cite{lzma} replace exact repeated strings, while block-sorting approaches such as Bzip2~\cite{bzip2} apply the Burrows-Wheeler Transform~\cite{adjeroh2008burrows} to create compressible runs of identical bytes. Statistical compressors like PPMd~\cite{cleary1984data} predict characters using probabilistic models based on context.

In contrast, log-specific compressors are tailored to log data's unique characteristics. As shown in Figure~\ref{fig:log-compression-paradigm}, they follow a multi-stage paradigm: logs are first segmented into chunks; each chunk undergoes preprocessing to identify repeated patterns (e.g., via parsing, rules, or similarity functions), and then are encoded into structured intermediate files, often through dictionary encoding and numeric encoding; finally, these files are compressed with a general-purpose tool (e.g., Gzip or LZMA) to achieve the final compression.

\subsection{Encoding Practice}
\subsubsection{Elastic Encoding}
Elastic encoding is a prominent variable-length integer encoding technique~\cite{wei2021feasibility,li2024logshrink,yu2024unlocking}.
It represents numbers using a sequence of bytes where the most significant bit (MSB) of each byte acts as a continuation flag.
An MSB of 1 indicates that the number continues in the next byte, while an MSB of 0 terminates the sequence.
For example, the integer \texttt{35} in a fixed-length 32-bit representation is \texttt{0000 0000 0000 0000 0000 0000 0010 0011}, occupying 4 bytes. 
With elastic encoding, it is compactly stored in a single byte \texttt{0010 0011}, where the MSB \texttt{0} terminates the sequence and the lower 7 bits store the value.
This approach greatly reduces integer storage requirements.
While effective, it is limited to purely numeric data and cannot compress the string-based or mixed-type tokens commonly found in logs.

\subsubsection{Delta Encoding}
Delta encoding is a common technique for compressing numeric data~\cite{wei2021feasibility, yu2024unlocking}.
It stores the difference (delta) between consecutive values instead of the absolute values, and encodes a base value followed by a sequence of deltas.
For example, the sequence \texttt{[100, 101, 102, 103, 104, 105]} is transformed by delta encoding into a base value of \texttt{100} and a delta sequence of \texttt{[1, 1, 1, 1, 1]}.
Since deltas are usually smaller, they lower entropy and improve compression.
Like elastic encoding, this method applies only to numeric data. Furthermore, it is only suitable for sequences with strong correlation (e.g., timestamps) and performs poorly on data without clear patterns (e.g., random port numbers or process IDs).

\subsubsection{Dictionary Encoding}
Dictionary encoding is a highly effective technique for compressing data with recurring values~\cite{yu2024unlocking,liu2019logzip}. It works by building a dictionary of unique data items and then replacing each occurrence of an item with a reference, or index, to its entry in the dictionary. 
While generic, dictionary encoding lacks flexibility. It treats each token as an opaque, monolithic unit, failing to capture intra-token patterns~(e.g., date tokens like \texttt{2015-07-29}).

\otherbox{
\textbf{Lesson Learned \#1:}
No single encoding strategy can effectively handle the variety of data in logs.
This highlights the need for a fine-grained hybrid encoding approach that combines tailored techniques for different data types.}

\begin{figure}[tbp]
    \centering
    \includegraphics[width=\linewidth]{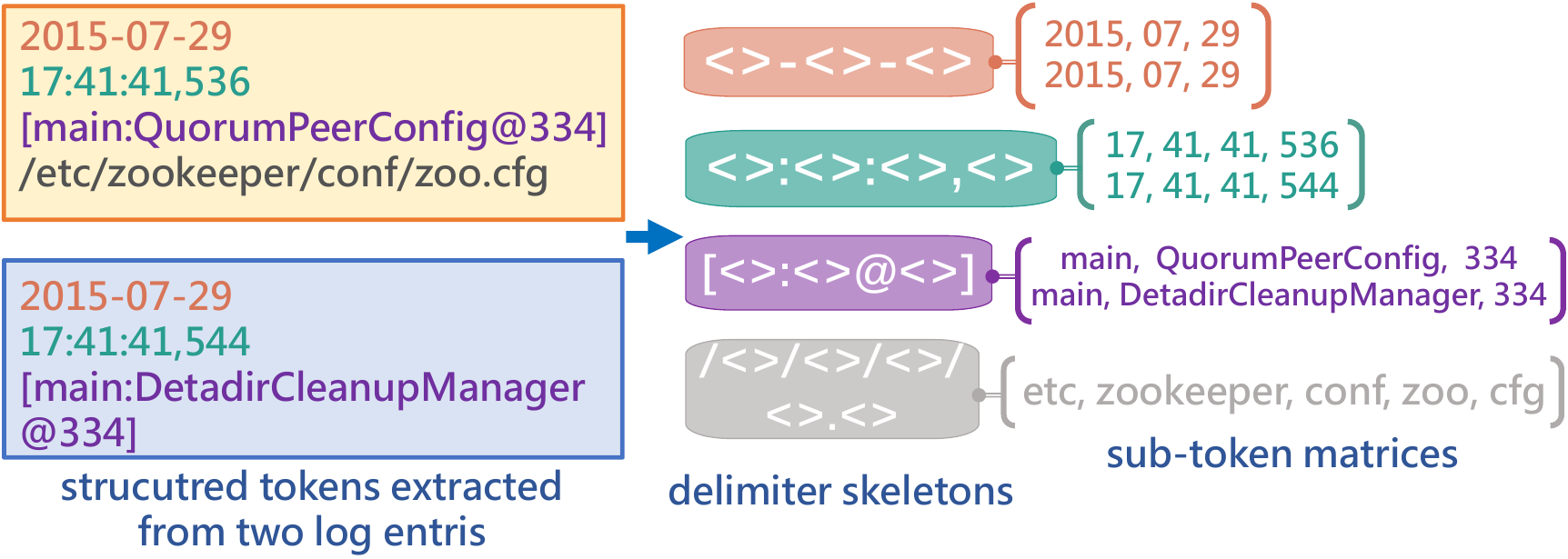}
            \vspace{-0.3in}
    \caption{Examples of structured tokens.}

    \label{fig:str-examples}
\end{figure}
\begin{figure*}[tbp]
    \centering
    \includegraphics[width=\linewidth]{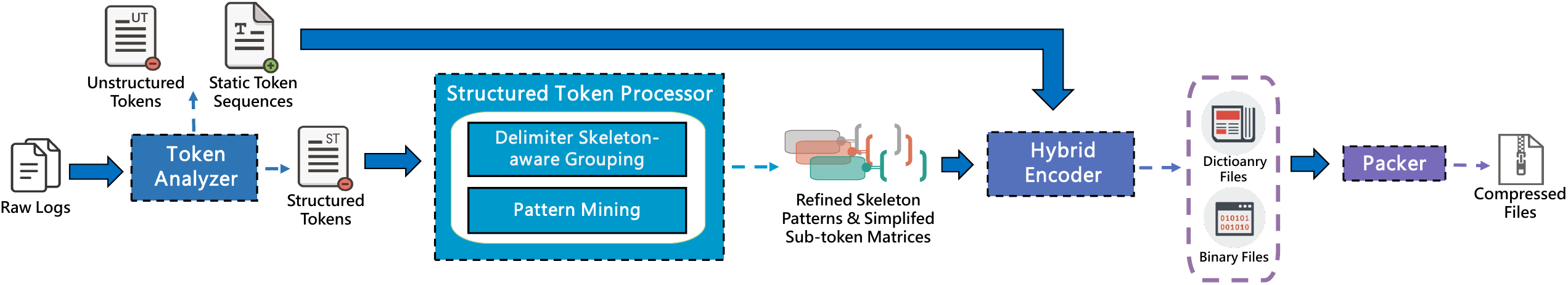}
    \vspace{-0.3in}
    \caption{Overview of \texttt{LogFold}.}
    \vspace{-0.1in}
    \label{fig:overview}
\end{figure*}
\subsection{Structured Tokens}
We identify a special type of \textit{structured token} common in system logs, which can be broken down into two main parts: a delimiter skeleton and a sub-token sequence. Although these complex tokens frequently occur, they have not been specifically addressed before.
The delimiter skeleton fixes the token's structural frame, made up of recurring non-alphanumeric characters{, as exemplified by Figure~\ref{fig:str-examples}.}
The sub-token sequence is the ordered list of variables that consists of numerals and letters. 
This decomposition reveals two additional redundancies for compression: 
(1) redundancy occurs when identical delimiter skeletons appear repeatedly across different tokens, and
(2) homogeneity exists within sub-tokens that include repeated or similar elements.
homogeneity.

\subsubsection{Delimiter Skeleton Redundancy}
We observe that a small, finite set of unique delimiter skeletons accounts for most structured token instances. 
For example, date tokens consistently follow a common skeleton pattern like \texttt{<>-<>-<>}, as illustrated in Figure~\ref{fig:str-examples}.
The frequent occurrence of these skeletons enables an effective strategy: replacing each unique skeleton with a compact identifier to eliminate repetitive patterns.

\subsubsection{Sub-token Homogeneity}
Structured tokens under the same delimiter skeleton can arrange their sub-token sequences into a conceptual matrix,
as shown in Figure~\ref{fig:str-examples}.
Sub-tokens in each matrix column correspond to the same structural position within the skeleton, resulting in homogeneous patterns such as data type, length, and value range.

This columnar homogeneity allows for tackling redundancy within each column and supports specialized encoding methods for the variable columns.
For example, in the sub-token matrix of time-based tokens (highlighted in green in the figure), the first three columns contain constant values, while the fourth exhibits a simple arithmetic sequence. This structure enables categorized processing: constants can be integrated into a refined skeleton with a more specific pattern (e.g., \texttt{17:41:41,<>}), while the arithmetic sequence can be compressed as dynamic information (e.g., delta encoding, elastic encoding).

\otherbox{
\textbf{Lesson Learned \#2:}
Breaking down structured tokens into delimiter skeletons and sub-token sequences: repeated skeletons and the homogeneity within sub-token columns. }

\section{Methodology}
We propose \texttt{LogFold}, an effective log compression method that leverages recurring patterns in structured tokens and employs a hybrid encoding strategy for enhanced compression.

The overall architecture of \texttt{LogFold} is depicted in Fig.~\ref{fig:overview}.
Taking raw logs as input, it outputs the compressed files through four components.
To begin with, the \textit{Token Analyzer} tokenizes each raw log entry and applies a set of predefined regular expressions to classify tokens as structured, unstructured, or static.
The static tokens from an entry constitute its static token sequence.
Next, the \textit{Structured Token Processor} handles the structured tokens in two phases.
Specifically, the \textit{Delimiter Skeleton-aware Grouping} phase groups structured tokens with identical delimiter skeletons to remove recurring patterns in these skeletons.
The \textit{Pattern Mining} phase further recognizes and eliminates recurring patterns within sub-tokens using pattern mining technologies.
Then the \textit{Hybrid Encoder} component takes the processed skeletons, sub-token matrices, unstructured tokens, and static token sequences, applying optimized hybrid encoding tailored for different data types, and outputs the intermediate files.
The \textit{Packer} then integrates these files and applies a general-purpose compressor to produce the final compressed output.

\subsection{Token Analyzer}
Unlike previous parsing-based approaches that separate logs by their template first~\cite{li2024logshrink, liu2019logzip}, we treat each log entry from a holistic view by partitioning it into its token types and preserving all header information. This allows us to align repetitive tokens across different headers and templates. For instance, a token like an IP address may act as an identifier in a header field and also appear in the message body describing a system event.
First, it tokenizes the raw entry and applies a fixed set of heuristic rules to distinguish dynamic tokens~(e.g., path, numerical values) from static tokens. 
Second, these dynamic tokens are further classified: tokens with an internal delimiter skeleton, {defined as an interwoven pattern of both alphanumeric and non-alphanumeric characters,} are labeled structured, while the rest are labeled unstructured. 
Finally, the analyzer replaces the original dynamic tokens with placeholders (<-> for structured, <*> for unstructured), leaving a sequence composed only of static tokens and placeholders.

\begin{figure}[tbp]
    \centering
    \includegraphics[width=\linewidth]{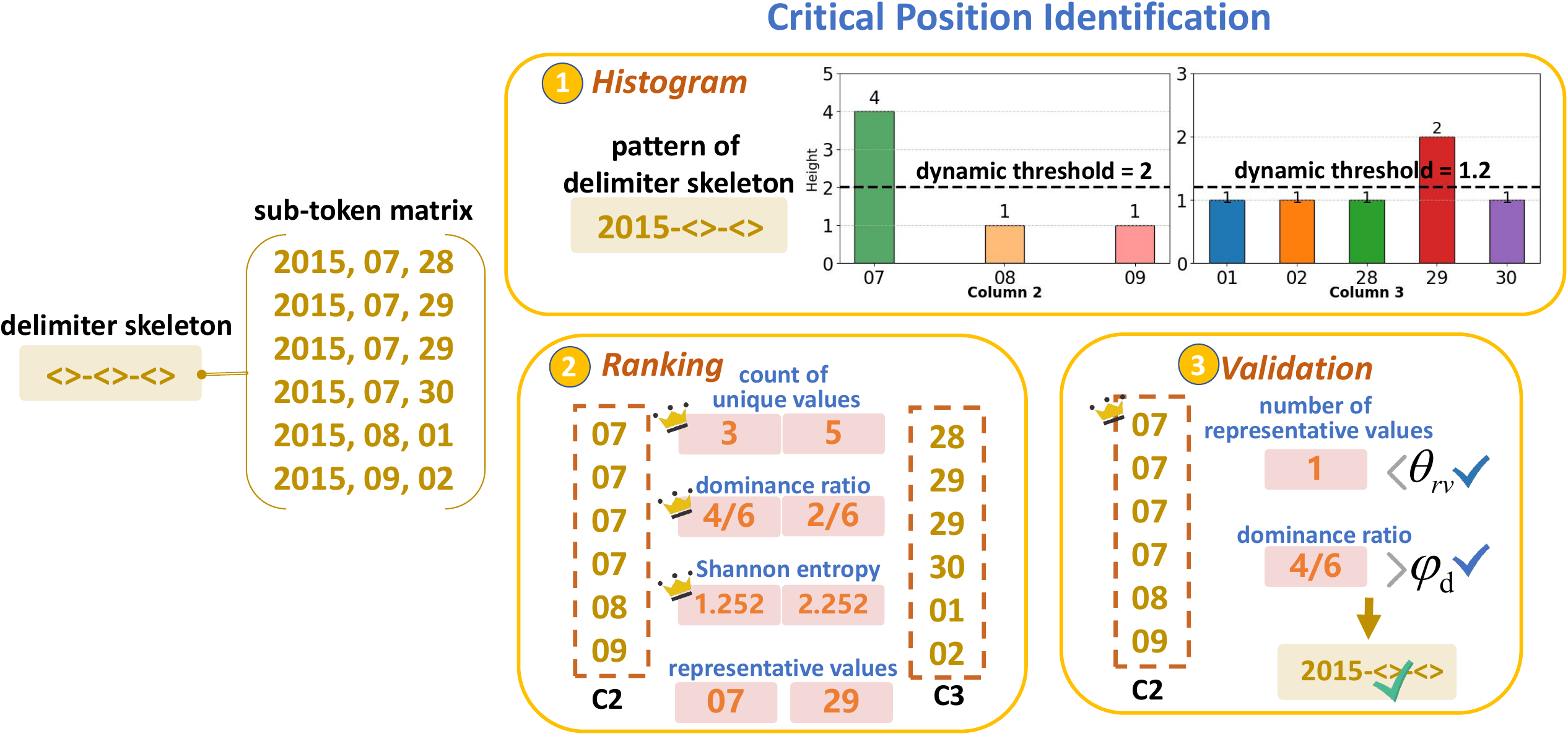}
    \caption{An example of the critical position identification step in the pattern mining phase.}
    \vspace{-0.1in}
    \label{fig:example-stp}
\end{figure}
\subsection{Structured Token Processor}
This component uses a two-phase process to recognize recurring patterns inside the structured tokens.

\subsubsection{Delimiter Skeleton-aware Grouping}
This phase clusters all structured tokens by their delimiter skeleton, then transforms their sub-tokens into a structural representation (see Fig~\ref{fig:str-examples}). This grouping mitigates skeleton redundancy and prepares the data for subsequent pattern mining. 
To achieve this, we scan each token character by character to extract its delimiter skeleton, which is composed of all non-alphanumeric characters. 
The skeleton then serves as the grouping key.
Consequently, each group is defined by a single, common delimiter skeleton and contains a corresponding matrix of sub-tokens.
\vspace{-0.1in}
\subsubsection{Pattern Mining}
The second phase comprises a three-step procedure to \textit{detect recurring patterns within sub-token matrices} (i.e., grouped sub-tokens).
The underlying principle is that, within a given delimiter skeleton, certain columns are often dominated by specific values, such as a common prefix for a system component or the same year in a date.

Motivated by this observation, we analyze the matrix vertically (column by column) to identify these special columns.
We define these columns as \textbf{critical positions} and their dominant values as \textbf{representative values}. 
By leveraging such position information, we partition the initial groups into smaller, more homogeneous sub-groups, thereby exposing deeper redundancies within sub-tokens.

\subsubsection*{\textbf{(1) Critical Position Identification.}}
The primary objective of this step is to identify the most critical position within a delimiter skeleton's sub-token matrix. 
This identification process comprises three steps: representative value identification, candidate ranking, and validation.

First, we identify representative values for each column using a histogram-based approach. 
A value is considered \textit{representative} if its frequency (i.e., the histogram bar height) meets or exceeds a dynamic threshold, calculated as the total number of rows divided by the count of unique values in that column. Taking the second column in Fig~\ref{fig:example-stp} as an example, the dynamic threshold is \texttt{\#rows(6)/\#uniq\_values(3)=2}. If a column has only one unique value, that value is directly included in the skeleton as a constant.
{This simple yet effective heuristic is based on the premise that dominant values should inherently appear more frequently.}

Second, {after all columns undergo this identification process,}
columns with at least one representative value are selected as the critical position candidates. 
{This phase aims to select the single most effective critical position for partitioning the matrix, as splitting by more than one column simultaneously would create overly sparse sub-matrices and thus harm compression effectiveness. To achieve this,} these candidates are ranked using a three-level heuristic that identifies the column with the lowest diversity and most uniformity, making it ideal for dividing data into homogeneous subgroups.
Specifically, the ranking first favors the column with the fewest unique values. 
If tied, it selects the column with the highest dominance ratio, defined as $\frac{N_{rv}}{N_{all}}$, where $N_{rv}$ is the occurrences of representative values and $N_{all}$ is the total number of rows. 
If there is any further draw, \texttt{LogFold} chooses the column with the lowest Shannon entropy~\cite{shannon1948mathematical}.
{This selection is guided by metrics that favor columns with less diversity (i.e., fewer unique values and lower Shannon entropy), as this characteristic helps prevent excessive fragmentation. Moreover, columns with a higher dominance ratio are prioritized because this indicates that a substantial number of rows will be grouped under the dominant value. This ensures the resulting sub-matrix, where the skeleton is refined, is sufficiently dense for effective compression.}

Third, the top candidate undergoes final validation to avoid overly specific, statistically insignificant patterns, maintaining robustness in our partitioning strategy. In particular, a critical position is accepted only if the number of representative values is below the threshold $\theta_{rv}$ or the dominance ratio exceeds $\varphi_{d}$. 
These criteria ensure partitioning occurs only when a dominant pattern is evident, preventing the formation of trivial, dispersed sub-token matrices that would undermine compression effectiveness\footnote{We investigate the impact of these thresholds in Sec~\ref{sec:sensitive-section}.}.

Fig.~\ref{fig:example-stp} illustrates the full process starting with tokens sharing the delimiter skeleton \texttt{<>-<>-<>}.
In this example, Column 1, containing only the value 2015, is immediately identified as constant, refining the skeleton from \texttt{<>-<>-<>} to the more specific \texttt{2015-<>-<>}, and removing C1 from further steps.
Next, for the remaining variable columns (C2 and C3), we identify representative values by building histograms for each column (Step {\textbf{\textcolor{log_example_blue}{\ding{202}}}}). In C2, the dynamic threshold is calculated as 2 (6 total entries / 3 unique values). Since the value \texttt{07} appears 4 times ($\ge 2$), it qualifies as a representative value. Similarly, for C3, \texttt{29} is a representative value.
With the identified representative values, we then rank the candidates to find the critical position (Step {\textbf{\textcolor{log_example_blue}{\ding{203}}}}): between C2 and C3, C2 is chosen due to fewer unique values (3 vs. 5). Finally, validation (Step {\textbf{\textcolor{log_example_blue}{\ding{204}}}}) compares C2's metrics against thresholds $\theta_{rv}$ and $\varphi_{d}$; if satisfied, the second placeholder is confirmed as a valid critical position.

\begin{figure}[tb]
    \centering
    \includegraphics[width=0.9\linewidth]{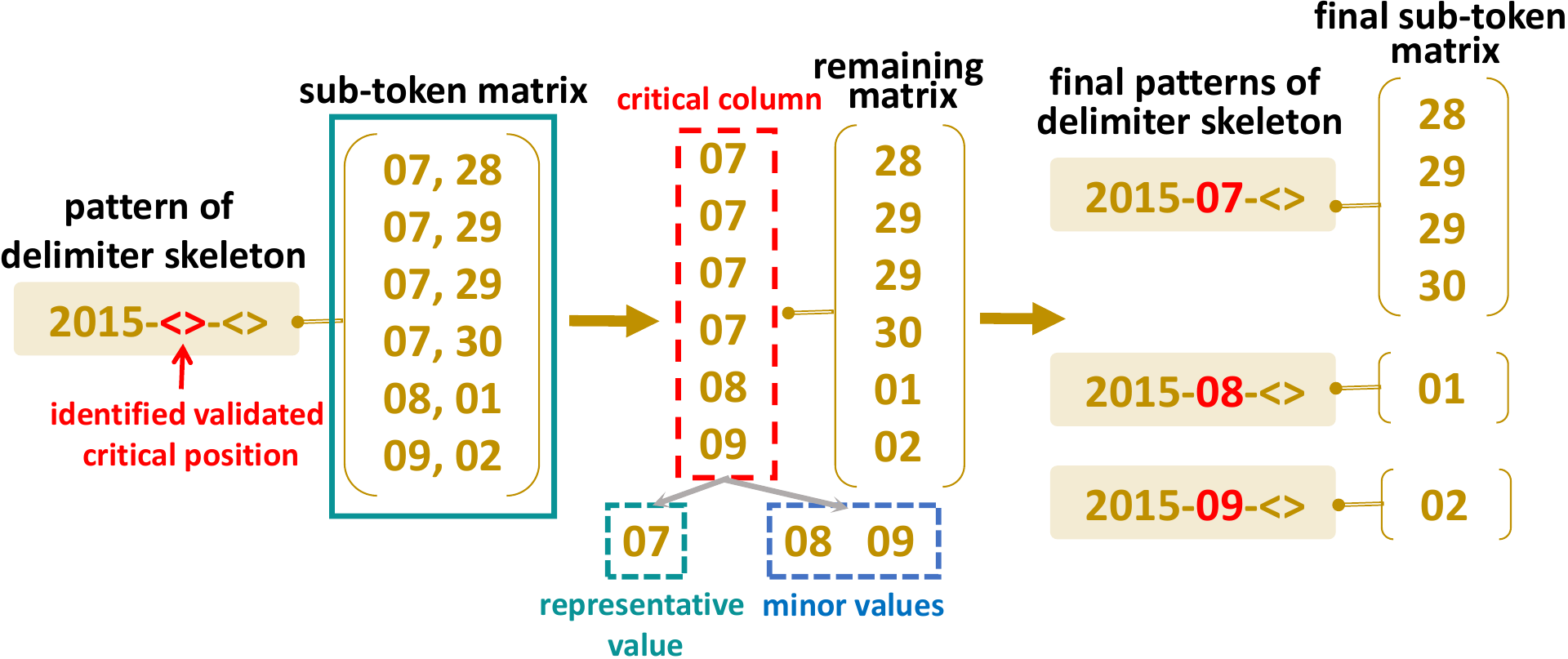}
    \vspace{-0.2in}
    \caption{An example of the re-grouping step.}
    \vspace{-0.2in}
    \label{fig:re-example}
\end{figure}
\subsubsection*{\textbf{(2) Re-grouping}}
Following the identification of a critical position, this step partitions the sub-token matrix into smaller sub-matrices~(i.e., subgroups). This further merges recurring patterns for each subgroup, leading to a more specific delimiter skeleton. Nevertheless, there is a trade-off: a more specific skeleton results in a sparse representation (e.g., one sub-matrix for each refined skeleton), while a too general skeleton fails to capture redundancy.

To address this, we employ a conditional re-grouping mechanism that balances specificity with pattern complexity.
Values in the critical column are categorized as \textit{representative} (identified previously) or \textit{minor} (all other unique values). 
This mechanism chooses between \textit{full} re-grouping and \textit{partial} re-grouping.
In specific, the full re-grouping strategy creates a distinct pattern for every unique value~(i.e., both representative and minor), by embedding each into the pattern's placeholder.
Partial re-grouping generates patterns only for representative values, retaining all minor-value rows under the original, more generic pattern. 
A threshold $\zeta_{uv}$ determines the strategy: full re-grouping is used when unique values are few (e.g., below $\le 3$) to maximize simplification; otherwise, partial re-grouping avoids overfitting caused by many sparse patterns.

Fig.~\ref{fig:re-example} illustrates the re-grouping process. 
The input sub-token matrix follows the skeleton~\texttt{2015-<>-<>}, with the second column identified as the critical position.
This column contains one representative value (i.e., \texttt{07}) and two minor values (i.e., \texttt{08}, \texttt{09}).
Because the number of minor values (2) is below the threshold (e.g., $\le 3$), the full re-grouping strategy is applied. 
This splits the original group into three specific subgroups with refined patterns \texttt{2015-07-<>}, \texttt{2015-08-<>}, and \texttt{2015-09-<>}. As a result, the resulting sub-token matrices become more homogeneous (in this case, reduced to a single column).

\subsubsection*{\textbf{(3) Final Pattern Refinement.}}
The final step further refines the skeleton patterns by identifying and embedding any new constant columns that appear after re-grouping.
After partitioning, some columns in the resulting sub-token matrices may become constant (i.e., containing a single unique value across all rows). To integrate these constants into the patterns, we apply FP-Growth~\cite{dur2000three, grahne2005fast, agrawal1993mining}, an efficient algorithm for mining frequent patterns in transactional data.
We treat each row as a transaction, with items represented as sub-token values paired with their column positions~(e.g., value \{col\_index\}). By setting the minimum support to 100\%, FP-Growth identifies only patterns present in every row, corresponding exactly to constant columns.
These constants are then embedded into the associated placeholders within the skeleton, yielding the final skeleton patterns.
Re-visiting the example in Figure~\ref{fig:re-example}, the sub-token matrices for patterns (\texttt{2015-08-<>} and \texttt{2015-09-<>}) reduce to single constants (\texttt{01} and \texttt{02}, respectively) after re-grouping. The algorithm detects and embeds these constants into the placeholders, yielding final patterns \texttt{2015-08-01} and \texttt{2015-09-02}. 

\subsection{Hybrid Encoder}
This component tailors optimized, column-wise encoding strategies to different types of data and thus effectively enhances the compression ratio.
Specifically, it uses three encoding strategies: 
(1) optimized numeric encoding,
(2) dictionary encoding and 
(3) mixed-type encoding.

\subsubsection{Optimized Numeric Encoding}
The encoding strategy is applied to the numbers~(i.e, the unstructured number tokens and sub-token matrices with only numbers).
We propose a \textit{dynamic delta encoding strategy} with two optimizations.

$\bullet$ \textit{Dynamic Delta Encoding.}
Following previous work~\cite{li2024logshrink,yu2024unlocking,wei2021feasibility}, we apply delta encoding to numerical columns for optimal effectiveness. 
It samples the first ten values of a column and compares the average of the absolute values for the original numbers and their delta-encoded forms.
If the average magnitude of the delta-encoded average is smaller, it indicates that the data changes gradually or contains some patterns. So we apply the delta encoding to the entire column. 
Otherwise, original values are kept to avoid the overhead of encoding to non-sequential or random data.

$\bullet$ \textit{Length-based Optimization.}
For purely numerical unstructured tokens, we take a step before applying the dynamic delta encoding.
The optimization first groups them by length, as length often reflects their value range. 
Each length-based group is then compressed with dynamic delta and elastic encoding strategies to generate compact binary representations. 
To preserve type consistency in the static token sequence, each length group is assigned a unique alphabetic tag~(e.g., tokens of length 1 are tagged \texttt{<a>}), which replaces the original placeholders in the static token sequences. 
These tags are then encoded together with the static token sequences using dictionary encoding.

\begin{figure}[t]
    \centering
    \includegraphics[width=0.95\linewidth]{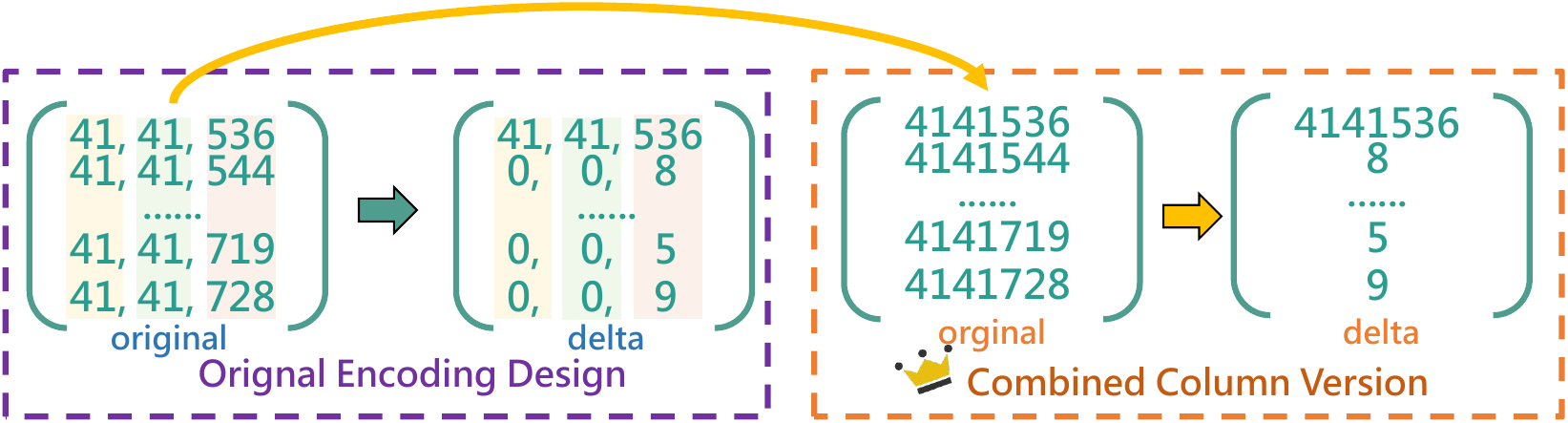}
    \caption{An example of the combined column optimization strategy.}
    \vspace{-0.2in}
    \label{fig:cc-example}
\end{figure}

$\bullet$ \textit{Combined Column Optimization.}
For matrices consisting only of numeric sub-tokens, this optimization aims to further reduce storage by identifying simpler arithmetic progressions within certain numeric matrices. It targets sub-token matrices representing compound entities (e.g., timestamps) composed entirely of fixed-length numbers. Instead of handling columns separately, this optimization concatenates the sub-tokens into a single large integer for each log entry~(e.g., sub-token sequence~\texttt{[2015, 07, 28]} becomes \texttt{20150728}). 
By applying delta encoding to the sequence of these large concatenated integers, we can capture the underlying progression more effectively than encoding deltas of individual columns. 
To apply this optimization only when beneficial, we use a lightweight heuristic: after sampling the data, if the concatenated delta sequence is more compact, then we adopt this strategy; otherwise, the system defaults to the standard dynamic delta encoding design.

Fig.~\ref{fig:cc-example} shows an example of the combined column optimization.
The original sub-token matrix appears on the left side, with each column storing the raw values. Delta encoding is then applied column-by-column, as indicated by the green arrow (e.g., the first column yields deltas \texttt{[41, 0,} ...,\texttt{ 0, 0]} derived from repeated values \texttt{[41, 41,} ...,\texttt{ 41, 41]}).
In this optimization strategy, we first concatenate all three columns into a single column of larger integers (e.g., \texttt{4141536} for the sequence \texttt{[41, 41, 536]}, highlighted in orange). We then compute the delta version of this concatenated column (shown on the far right).
Comparing the delta sequences from the original encoding (multi-column) and the combined-column version (which reduces three columns to one), the latter proves more compact. Consequently, the optimization is adopted, and we apply elastic encoding to the single column for space savings.

\subsubsection{Dictionary Encoding}
For unstructured tokens, including only strings, skeleton patterns, and the static token sequences, we use the traditional dictionary encoding strategy.
We first use a token dictionary to store unstructured strings and the skeleton patterns.
To clearly distinguish data sources, placeholders for structured tokens (i.e. \texttt{<->}) are replaced with a \texttt{|tag|} format, while unstructured strings retain the \texttt{<*>} placeholder.
Then, we use a separate template dictionary to store the static token sequences containing various tags and placeholders:~(i.e., \texttt{|tag|} for structured tokens' delimiter patterns, \texttt{<length-based tag>} for unstructured numbers, and \texttt{<*>} for unstructured strings).
The original order of IDs, corresponding to log entries and unstructured strings, is preserved and encoded using elastic encoding, respectively.
This process results in two dictionary files and a set of binary ID data files.

\begin{figure}
    \centering
    \vspace{-0.1in}
\includegraphics[width=0.9\linewidth]{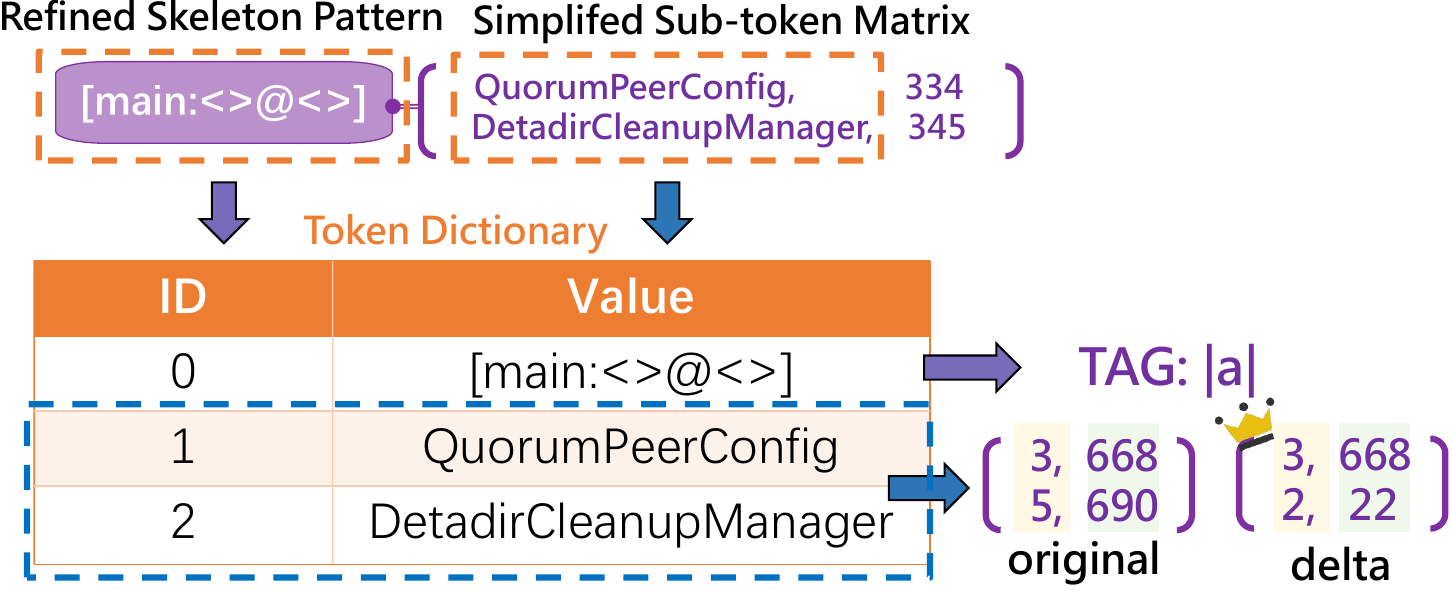}
    \caption{An example of the dictionary-based, mixed-type, and dynamic delta encoding.}
    \vspace{-0.2in}
    \label{fig:mix-example}
\end{figure}

\subsubsection{{Mixed-Type Encoding}}
This encoding strategy enables uniform numerical encoding of sub-token matrices that contain \textit{both numerical and string values}, thereby further exploiting the benefits of numeric encoding schemes. 
First, all string values are mapped to unique numerical IDs,  stored in a dictionary for later reconstruction.
Leveraging columnar homogeneity (i.e., sub-tokens within a column likely share the same type), we assign contiguous IDs to strings within each column. To distinguish original numbers from these string-derived IDs in a unified numerical space and ensure lossless compression, we apply a parity-based transformation: original numbers $n$ become \textit{even} integers ($2*n$), while string IDs $id$ become \textit{odd} integers ($2*id + 1$).
This allows the entire column to be uniformly encoded (e.g., by elastic encoding or dynamic delta encoding), thereby maximizing compression effectiveness while guaranteeing a lossless and reversible transformation.

Figure~\ref{fig:mix-example} illustrates the mixed-type encoding strategy.
The simplified sub-token matrix contains a mix of numeric and string values~(e.g., \texttt{QuorumPeerConfig} with number \texttt{344}). To enable uniform numerical encoding, strings are first assigned unique, contiguous numeric IDs on a column-by-column basis (e.g., strings in the first column receive IDs \texttt{1} and \texttt{2}). These IDs replace the original strings, while a dictionary records the mappings for reconstruction.
To distinguish original numbers from string-derived IDs and ensure lossless compression, a parity transformation is applied, with numbers mapped to their even integers~(e.g., \texttt{334} becomes \texttt{668}) and IDs to their odd integers~(e.g, \texttt{1} becomes \texttt{3}).
This yields a fully numerically encoded matrix~(shown as the original version).
Finally, dynamic delta encoding is applied to this matrix~(e.g, the second column \texttt{[668, 690]} to deltas like \texttt{[668, 22]}), resulting in a more compact representation that leverages arithmetic progressions for enhanced compression.

\subsection{Packer}
The \textit{Packer} module, like other log compressors, is the final step in \texttt{LogFold}. It aggregates all intermediate outputs from earlier components~(i.e., the dictionary and all binary data files) into one archive. This archive then goes through a general-purpose compressor (e.g., Gzip) to produce the final compressed file.

\subsection{Decompressor}
\texttt{LogFold} provides lossless decompression by reversing its compression pipeline. 
It guarantees that no information is lost when compressing the log data.
The process starts with decompressing the main archive. 
Next, it decompresses the ID binary file of static token sequences. 
With these decoded IDs, along with the template dictionary, the original static token sequences are recovered.
Then, \texttt{LogFold} decodes the binary files of the number tokens and replaces the corresponding tags in the ordered static token sequences with these decoded values.
Subsequently, it decodes the ID binary files of strings and the matrix binary files. 
Finally, it uses the token dictionary to map these decoded matrix values back to the ordered static token sequences to restore the original log data.

\section{Implementation}
We implement \texttt{LogFold} in C++, consisting of approximately 2,600 lines of code measured by cloc~\cite{cloc}. 
For our experiments, \texttt{LogFold} was configured with these default settings: 
the \textit{Packer} component utilizes \textit{tar} for archiving and \textit{LZMA} for final compression. The key thresholds for critical position identification were set to $\theta_{rv}=40$ and $\varphi_{d}=0.6$, and the threshold for re-grouping strategy decision was set to $\zeta_{nv}=3$.
{We use regular expression
\texttt{(([\textbackslash p\{L\}\textbackslash p\{N\}]+)|([\^{}\textbackslash p\{L\}\textbackslash p\{N\}]+))} to identify delimiters.}
The default strategy for token analyzer combines numeric and path expressions, using the regular expressions \texttt{\^{}\textbackslash S*\textbackslash d\textbackslash S*\$} and \\ \texttt{(/[\^{}/]*)+|([a-zA-Z]:\textbackslash\textbackslash(?:[\^{}\textbackslash\textbackslash]*\textbackslash\textbackslash)}, respectively.

\section{Evaluation}
\begin{table}[!tbp]
\footnotesize
\caption{The statistics of the public log datasets.}
\vspace{-0.1in}
\label{tab:sorted-dataset-sizes}
\begin{tabular}{l l l l}
\toprule
\textbf{System Type} & \textbf{Dataset} & \textbf{File Size} & \textbf{\# Lines} \\
\toprule
\multirow{3}{*}{Supercomputers} 
    & Thunderbird & 29.60 GB & 211,212,192 \\
    & BGL & 708.76 MB & 4,747,963 \\
    & HPC & 32.00 MB & 433,489 \\
\addlinespace
Operating systems 
    & Windows & 26.09 GB & 114,608,388 \\
    & Mac & 16.09 MB & 117,283 \\
    & Linux & 2.25 MB & 25,567 \\
\addlinespace
\multirow{2}{*}{Distributed systems} 
    & Spark & 2.75 GB & 33,236,604 \\
    & HDFS & 1.47 GB & 11,175,629 \\
    & OpenStack & 60.01 MB & 207,820 \\
    & Hadoop & 48.61 MB & 394,308 \\
    & Zookeeper & 9.95 MB & 74,380 \\
\addlinespace
Mobile systems 
    & Android & 183.37 MB & 1,555,005 \\
    & HealthApp & 22.44 MB & 253,395 \\
\addlinespace
Server applications 
    & OpenSSH & 70.02 MB & 655,146 \\
    & Apache & 4.96 MB & 56,481 \\
\addlinespace
Standalone software 
    & Proxifier & 2.42 MB & 21,329 \\
\bottomrule
\end{tabular}
\end{table}
We conduct evaluations to answer the following research questions:

\textbf{RQ1: How well does \texttt{LogFold} improve log compression?}

\textbf{RQ2: How do different components contribute to \texttt{LogFold}'s effectiveness?}

\textbf{RQ3: How sensitive is \texttt{LogFold} to its internal parameter settings?}

\textbf{RQ4: How generalizable is \texttt{LogFold} across
different zip tools {with different compression levels}?}

{\textbf{RQ5: How does \texttt{LogFold} perform in log decompression?}}

Consistent with prior work~\cite{li2024logshrink,yu2024unlocking}, the input log data is processed in chunks of 100,000 lines to facilitate scalable processing.
All experiments were conducted within a Docker container running Ubuntu 22.04, hosted on an x86 64-bit server equipped with an 80-core Intel Xeon Gold 5218R CPU @ 2.10GHz, 267 GiB of RAM, and a 5.6 TB disk.

\subsubsection{Log Dataset}
Our evaluation is conducted on a benchmark of 16 publicly available log datasets~\cite{jiang2024large,zhu2023loghub}, selected to be consistent with prior studies~\cite{li2024logshrink,yu2024unlocking}. 
These datasets originate from a wide array of systems, ranging from supercomputers and distributed systems to standalone software, and collectively amount to over 77 GB in size. 
This diversity in source, scale, and structure provides a robust foundation for a comprehensive evaluation of \texttt{LogFold}'s performance and generalizability. 
Detailed statistics for each dataset are summarized in Table~\ref{tab:sorted-dataset-sizes}.

\subsubsection{Evaluation Metrics}
To evaluate the performance of \texttt{LogFold}, we employ two standard metrics, consistent with prior work in log compression~\cite{li2024logshrink,liu2019logzip,yu2024unlocking}. First, to measure effectiveness, we use the Compression Ratio~(CR), which quantifies the degree of data reduction. A higher CR indicates more effective compression. Second, to measure efficiency, we use Compression Speed~(CS), which reflects the throughput of the entire process. The metrics are formally defined as follows:
$
CR = \frac{Original\ Size}{Compressed \ Size} 
$
and
$
CS = \frac{Original\ Size}{Total\ Processing\ Time}
$ where $Total\ Processing\ Time$ encompasses the full duration required to produce the final compressed files.
{We also include decompression speed for evaluations in log decompression. The metric is defined as: $DS = \frac{Original \ Size}{Decompression \ Time}$ following former work~\cite{yao2020study}}.
\subsubsection{Baselines}
Following prior work~\cite{yu2024unlocking}, we compare \texttt{LogFold} with {nine}
baselines. 
These include four general-purpose compressors (gzip, LZMA, bzip2, and PPMd), which operate on raw byte streams using established dictionary-based, block-sorting, and statistical techniques, respectively.
More importantly, we benchmark against {five}
state-of-the-art log compressors. 
{LogZip}\cite{liu2019logzip} exemplifies the classic parser-based approach of separating templates from parameters.  
{LogReducer}\cite{wei2021feasibility} and {LogShrink}\cite{li2024logshrink} build upon this paradigm by applying advanced optimizations to numerical data and patterns. 
{Similarly, LogBlock~\cite{yao2021improving} identifies recurring patterns by log parsing and employs heuristics to preprocess small log blocks, reducing log repetitiveness and rearranging content prior to general compression.}
In contrast, {Denum}\cite{yu2024unlocking} offers a non-parsing alternative, distinguishing itself by first extracting and compressing numeric tokens from the raw log stream. This selection of baselines provides a comprehensive benchmark against both general-purpose tools and the primary approaches in specialized log compression.
{Apart from the selected baselines, there are other log compressors, such as CLP~\cite{rodrigues2021clp}, LogGrep~\cite{wei2023loggrep}, Cowic~\cite{lin2015cowic}, and LogArchive~\cite{christensen2013adaptive}. We exclude CLP and LogGrep because they target at querying the compressed log data. Similarly, Cowic is designed to prioritize decompression speed over compression ratio. We exclude LogArchive because it has been shown to perform significantly worse than LogReducer. Consequently, they were not included in the evaluation.} 

For our evaluation, we adopt the compression ratios reported in the original papers~\cite{li2024logshrink,yu2024unlocking}, as this metric is environment-independent. 
For compression speed, which is hardware-sensitive, we employ the open-source implementations of all tools~\cite{LogReducer-code,Denum-code,LogShrink-code} in our controlled experimental environment to ensure a fair comparison.
Note that due to the poor performance of LogZip reported  in~\cite{li2024logshrink}, we directly use the reported performance metrics from~\cite{li2024logshrink}.
{In addition, the compression levels for each baseline compressor are their default setting for fair compression. For LogBlock, targeted at compressing small blocks of log data, we choose 128k and LZMA with default compression level~(i.e., Level 6) as its setting to trade-off CR and CS.}

\begin{table*}[htb]
    \caption{Statistics of compression ratio among compressors.}
    \vspace{-0.1in}
    \label{tab:rq1-result}
    \footnotesize
    \begin{tabular}{l *{10}{c}} 
    \toprule
        Dataset & gzip & LZMA & bzip2 & PPMd & LogZip & \textcolor{black}{LogBlock}& LogReducer & LogShrink & Denum  & LogFold \\ 
        \toprule
        Android & 7.742 & 18.857 & 12.787 & 19.370 & 25.165 &\textcolor{black}{22.666} & 20.776 & 21.857 & 32.494 & \textbf{32.716}\\ 
        Apache & 21.308 & 25.186 & 29.557 & 31.688 & 30.375 &\textcolor{black}{33.067}& 43.028 & 55.940 & \textbf{58.517} & 57.809 \\ 
        BGL & 12.927 & 17.637 & 15.461 & 18.927 & 32.655 &\textcolor{black}{36.421} & 38.600 & 42.385 & 41.804 & \textbf{51.019} \\ 
        Hadoop & 20.485 & 36.095 & 32.598 & 32.110 & 35.008 &\textcolor{black}{59.566} & 52.830 & 60.091 & 78.546 & \textbf{90.971} \\ 
        HDFS & 10.636 & 13.559 & 14.059 & 19.155 & 16.666 &\textcolor{black}{18.440}& 22.634 & \textbf{27.319} & 25.670 & 25.217 \\ 
        HealthApp & 10.957 & 13.431 & 13.843 & 15.337 & 22.634 &\textcolor{black}{15.803}& 31.694 & 39.072 & 44.472 & \textbf{56.562} \\ 
        HPC & 11.263 & 15.076 & 12.756 & 14.822 & 27.208 &\textcolor{black}{32.000}& 32.070 & 35.878 & 45.275 & \textbf{46.950} \\ 
        Linux & 11.232 & 16.677 & 14.695 & 18.508 & 23.368&\textcolor{black}{25.000} & 25.213 & 29.252 & 30.449 & \textbf{31.508} \\ 
        Mac & 11.733 & 22.159 & 18.074 & 28.469 & 26.306 &\textcolor{black}{28.228}& 35.251 & 39.860 & 40.789 & \textbf{49.399} \\ 
        OpenSSH & 16.828 & 18.918 & 22.865 & 31.977 & 42.606 &\textcolor{black}{49.660}& 86.699 & \textbf{103.175} & 101.654 & 96.758 \\ 
        OpenStack & 12.158 & 14.437 & 15.231 & 17.429 & 17.258 &\textcolor{black}{17.650} & 16.701 & 22.157 & \textbf{22.238} & 20.390 \\ 
        Proxifier & 15.716 & 18.982 & 23.619 & 25.489 & 21.493 &\textcolor{black}{22.000}& 25.501 & 27.029 & 27.288 & \textbf{29.143} \\ 
        Spark & 17.825 & 19.908 & 26.497 & 30.614 & 20.825&\textcolor{black}{27.095} & 59.470 & 59.739 & 59.470 & \textbf{59.881} \\ 
        Thunderbird & 16.462 & 27.309 & 25.428 & 33.026 & - &\textcolor{black}{40.068}& 49.185 & 48.434 & 63.824 & \textbf{67.686} \\ 
        Windows & 17.798 & 202.568 & 67.533 & 61.083 & 310.596 &\textcolor{black}{238.729} & 342.975 & 456.301 & 481.350 & \textbf{570.552} \\ 
        Zookeeper & 25.979 & 27.667 & 36.156 & 38.931 & 47.373 &\textcolor{black}{55.278} & 94.562 & 116.981 & 135.251 & \textbf{145.741} \\ 
        \bottomrule
        \midrule

        Average & 15.066 & 31.779 & 23.822  & 27.308 & 46.636 &\textcolor{black}{45.042}& 61.074 & 74.092 & 80.568 & \textbf{89.519} \\ 

        \bottomrule 
    \end{tabular}
    \vspace{-0.1in}
\end{table*}

\begin{table}[tbp]
    \footnotesize
    \caption{Compression speed~(MB/s) among log compressors.}
    \vspace{-0.1in}
    \label{tab:rq1-cs}
    \centering
    \begin{tabular}{l c c c c c c}
    \toprule
    \multirow{1}{*}{Dataset} & 
    \multicolumn{1}{c}{\multirow{1}{*}{LogZip}} & 
    \multicolumn{1}{c}{\multirow{1}{*}{{\textcolor{black}{LogBlock}}}} & 
    \multicolumn{1}{c}{\multirow{1}{*}{LogReducer}} & 
    \multicolumn{1}{c}{\multirow{1}{*}{LogShrink}} & 
    \multicolumn{1}{c}{Denum} & 
    \multicolumn{1}{c}{\multirow{1}{*}{LogFold}} \\
    \toprule
        Android & 0.068 & \textcolor{black}{11.510} & 11.668 & 0.870 & 32.193 & 6.066 \\ 
        Apache & 0.074 & \textcolor{black}{8.005} & 1.518 & 0.749 & 3.766 & 5.662 \\ 
        BGL & 0.874 & \textcolor{black}{12.188} & 13.339 & 1.629 & 40.896 & 13.340 \\ 
        Hadoop & 0.901 & \textcolor{black}{15.503} & 8.959 & 1.228 & 13.857 & 12.232 \\ 
        HDFS & 0.701 & \textcolor{black}{3.929} &14.416 & 1.629 & 73.143 & 18.645 \\ 
        HealthApp & 0.736 & \textcolor{black}{3.078} &3.409 & 2.349 & 7.167 & 11.514 \\ 
        HPC & 0.644 & \textcolor{black}{7.738} &5.509 & 1.767 & 18.038 & 13.634 \\ 
        Linux & 0.687 & \textcolor{black}{4.149} &0.775 & 0.392 & 3.617 & 2.616 \\ 
        Mac & 0.009 & \textcolor{black}{7.088} &3.245 & 0.971 & 3.515 & 5.505 \\ 
        OpenSSH & 0.751 & \textcolor{black}{16.432} &8.875 & 2.034 & 33.908 & 15.703 \\ 
        OpenStack & 0.537 & \textcolor{black}{8.291} &8.525 & 1.823 & 10.541 & 9.916 \\ 
        Proxifier & 0.716 & \textcolor{black}{5.969} &0.848 & 0.351 & 2.827 & 3.523 \\ 
        Spark & 0.550 & \textcolor{black}{8.203} &16.022 & 1.145 & 86.129 & 16.662 \\ 
        Thunderbird & --- & \textcolor{black}{13.054} &9.561 & 2.702 & 106.009 & 5.922 \\ 
        Windows & 1.357 & \textcolor{black}{18.946} &6.823 & 4.868 & 85.935 & 10.987 \\ 
        Zookeeper & 0.842 & \textcolor{black}{10.098} &2.796 & 1.194 & 3.718 & 5.549 \\ 
    \bottomrule
    \midrule
    Average & 0.630 & \textcolor{black}{9.636} & 7.268 & 1.606 & 32.829 & 9.842 \\
    \bottomrule
    \end{tabular}
\end{table}
\subsection{RQ1: How well does \texttt{LogFold} improve log compression?}
Table~\ref{tab:rq1-result} and Table~\ref{tab:rq1-cs} show the results.
\texttt{LogFold} exhibits the highest average CR among all evaluated compressors, reaching a mean value of 89.519. This represents a 494.19\% improvement relative to the widely-used baseline compressor, gzip.
When compared with its peers, \texttt{LogFold} consistently demonstrates superior performance. It surpasses the second-most effective compressor, Denum (average ratio of 80.568), by a notable margin of approximately 11 points on the ratio scale. The robustness of \texttt{LogFold} is further demonstrated by its consistent performance on individual datasets. It secures the top CR in 12 out of 16 cases.
The exceptions arise on datasets with a higher proportion of unstructured strings~(e.g., HDFS), where LogFold defaults to a classic dictionary-based encoding, thereby limiting its compression ratio. 
It excels on diverse log types, including large-scale system logs (e.g., Windows), application logs (e.g., HealthApp), and distributed system logs (e.g., Hadoop, Zookeeper). This broad success underscores its wide applicability.

Regarding CS, \texttt{LogFold} operates at an average of 9.842 MB/s\textcolor{black}{, and LogBlock operates at a similar average speed of 9.636 MB/s.}
This performance is a direct result of its advanced compression strategy. Achieving a high CR inherently requires more sophisticated and computationally intensive algorithms for processing log data. Therefore, the processing speed represents a trade-off for its superior storage savings.
While other C++ based tools like Denum offer higher throughput, they do so with a lower CR. 
This positions \texttt{LogFold} as a tool specifically engineered for scenarios where the primary goal is maximizing data reduction, such as in long-term archiving or offline analysis, where processing time is less critical than minimizing storage footprint.
\answerbox{
\textbf{Answer to RQ1:}
\texttt{LogFold} establishes a new state-of-the-art in compression effectiveness with an average ratio of 89.519, outperforming all baselines. It achieves the highest compression ratios on 12 of 16 datasets, delivering an 11.11\% to 494.19\% improvement over competing methods with a practical compression speed.
}

\subsection{RQ2: How do different components contribute to \texttt{LogFold}'s effectiveness?}
To address this research question, we conducted a comprehensive ablation study by evaluating two variants of \texttt{LogFold}. 
The first variant, "w/o processor w/o encoder", disables both the Structured Token Processor and the Hybrid Encoder, handling all tokens (structured and unstructured) via conventional dictionary-based encoding. 
The second variant, "w/o processor", omits only the Structured Token Processor, passing the structured tokens directly to the Hybrid Encoder without preprocessing. Dictionary-based encoding serves as a well-justified baseline, representing a classic and widely adopted approach in prominent log compression tools~\cite{liu2019logzip, li2024logshrink,rodrigues2021clp,yu2024unlocking}.

\begin{figure}[!tbp]
    \centering
    \includegraphics[width=0.9\linewidth]{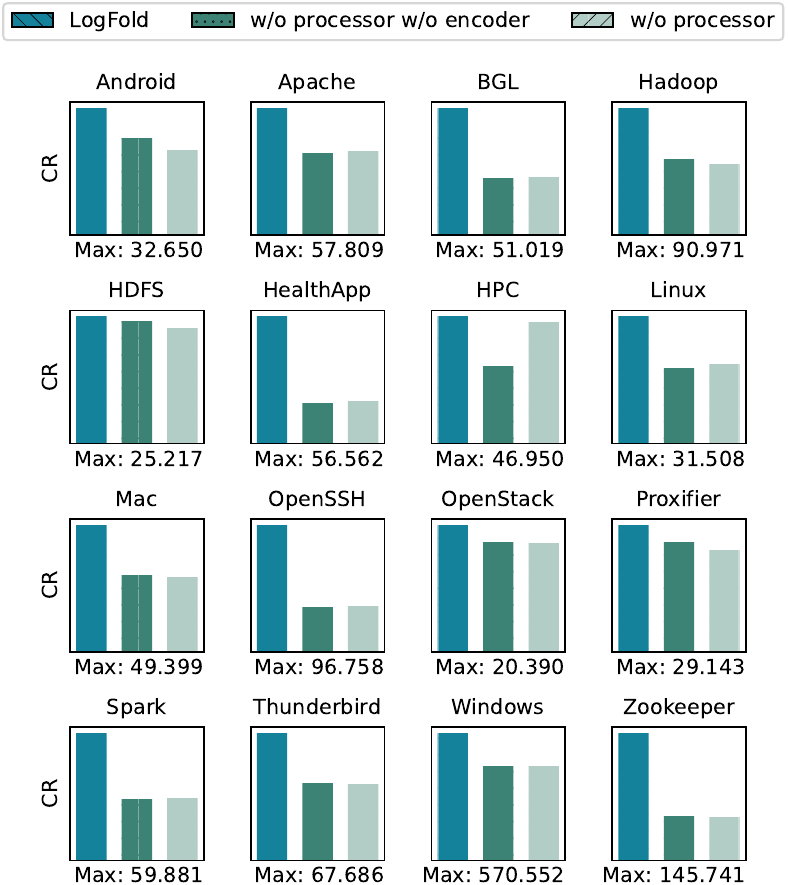}
    \vspace{-5pt}
    \caption{The results of the ablation study. In all 16 datasets, the full \texttt{LogFold} achieves the maximum CR, denoted by the value under each corresponding plot.}
    \label{rq2-results}
\end{figure}

Figure~\ref{rq2-results} presents the results of our ablation study, confirming that the complete \texttt{LogFold} implementation consistently achieves the highest CR across all 16 datasets. 
This study validates the significant contributions of both the {Structured Token Processor} and the {Hybrid Encoder}. The removal of the Structured Token Processor (i.e., "w/o processor" variant) always results in a significant performance degradation compared to the full \texttt{LogFold}, highlighting the processor's essential role. Similarly, the effectiveness of our Hybrid Encoder is demonstrated as the "w/o processor" variant generally outperforms the "w/o processor w/o encoder" variant, which relies on a simple dictionary encoder. Although the performance difference between the two variants fluctuates depending on the dataset characteristics, the overarching conclusion is that optimal results are only achieved when both components are utilized. Therefore, the results affirm that both the processor and the encoder are crucial and effective elements of \texttt{LogFold}'s design.

\answerbox{
\textbf{Answer to RQ2:} 
The Structured Token Processor and the Hybrid Encoder are both confirmed to be indispensable to \texttt{LogFold}, as the removal of either one consistently results in a substantial drop in compression ratio.
}

\subsection{RQ3: How sensitive is \texttt{LogFold} to its internal parameter settings?}\label{sec:sensitive-section}
\begin{figure}[tbp]
    \centering
    \begin{minipage}[b]{0.48\linewidth}
        \centering
        \includegraphics[width=\linewidth]{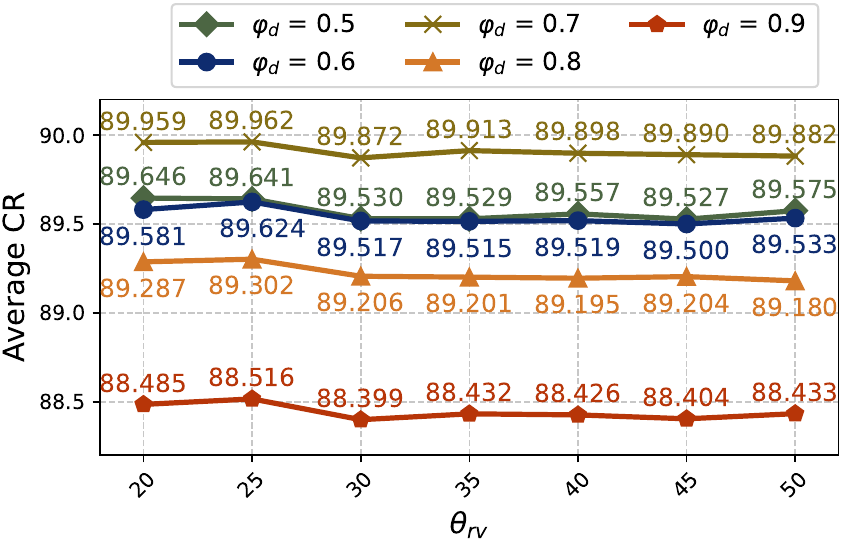}
        \caption{Average CR with different setting combinations of $\theta_{rv}$ and $\varphi_{d}$.} 
        \label{fig:rq3-1}
    \end{minipage}
    \hfill 
    \begin{minipage}[b]{0.48\linewidth}
        \centering
        \includegraphics[width=\linewidth]{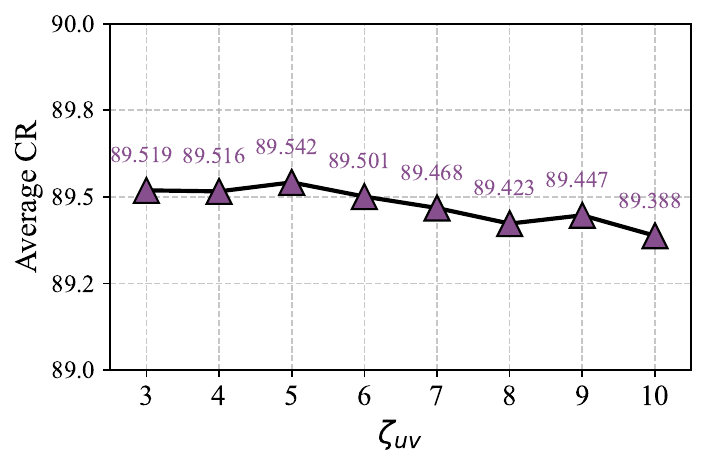}
        \caption{Average CR with different values of the threshold $\zeta_{uv}$.} 
        \label{fig:res-regroup}
    \end{minipage}
    \vspace{-10pt}
\end{figure}

\begin{figure}[tbp]
    \centering
    \includegraphics[width=0.7\linewidth]{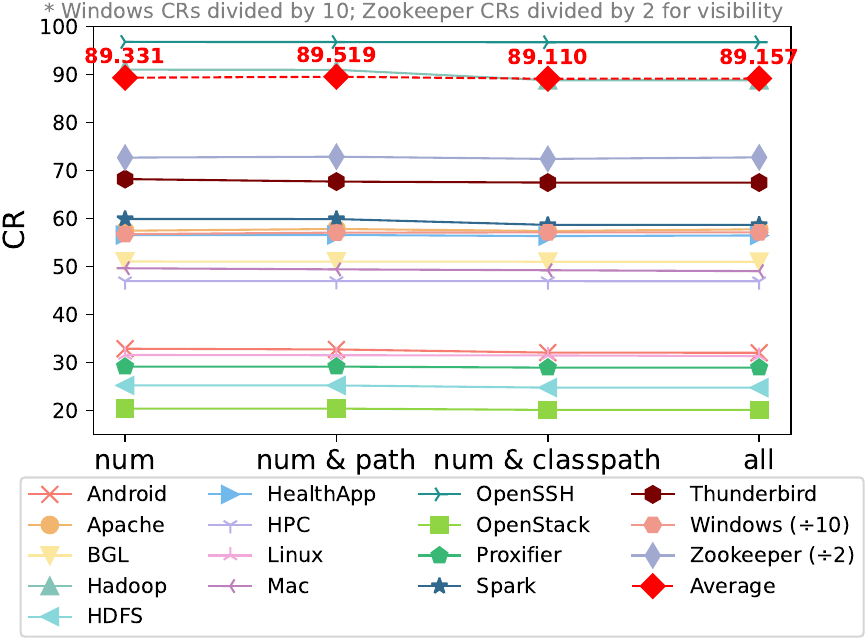}
    \vspace{-10pt}
    \caption{CR with different token selection strategies across 16 datasets.}
    \label{fig:rq3-2}
    \vspace{-15pt}
\end{figure}
\subsubsection{Critical Position Threshold Settings}
We investigate the sensitivity of \texttt{LogFold} to two key thresholds used in the critical position identification step of the Pattern Mining Phase.
The threshold $\theta_{rv}$ defines the maximum number of representative values, and the threshold $\varphi_{d}$ represents the required dominance ratio.
We varied $\theta_{rv}$ from 20 to 50 and $\varphi_{d}$ from 0.5 to 0.9 to explore the parameter space.
The results shown in Figure~\ref{fig:rq3-1} reveal that \texttt{LogFold} is highly robust against various parameter settings. In all combination settings, it consistently achieves average CR values between 88 and 90, significantly outperforming all existing baselines. 
First, its performance remains stable regardless of the  $\theta_{rv}$ value across the tested range. 
Second, although $\varphi_{d}=0.7$ achieves a slightly higher peak average CR, the difference between $\varphi_{d}=0.5$, $\varphi_{d}=0.6$, and $\varphi_{d}=0.7$ is marginal. 
We selected $\theta_{rv}=40$ and $\varphi_{d}=0.6$ as the default settings for \texttt{LogFold} for the reason of stability and conservatism. While $\varphi_{d}=0.7$ might offer a minor advantage on the current dataset, we believe the $\varphi_{d}=0.6$ setting offers more stability against variations in diverse log data. It may also more reliably reflect performance in multiple data sources in real-world practice.

\subsubsection{Re-grouping Threshold Settings.}
We investigate the sensitivity of \texttt{LogFold} to the threshold $\zeta_{uv}$, which determines whether to apply a full or partial re-grouping strategy. 
A larger threshold allows for more aggressive segmentation, but this could generate an excessive number of patterns and hinder overall compression. We therefore explore $\zeta_{uv}$ values from 3 to 10.
In the experiment, the average CR varies by only about 0.15 across the entire tested range, demonstrating that our method is robust to this parameter. 
As shown in Figure~\ref{fig:res-regroup}, performance peaks at $\zeta_{uv}=5$
 (i.e., 89.541) and then slightly declines, suggesting that overly liberal full re-grouping leads to pattern overfitting.
We chose $\zeta_{uv} = 3$ as a default, as it achieves a CR nearly identical to the peak while conservatively avoiding the risk of excessive pattern creation.

\subsubsection{Different Selection Strategies of Dynamic Tokens}
We evaluate four distinct selection strategies for dynamic tokens, which are constructed from three token types identified via source code analysis: numeric, path, and classpath tokens\footnote{regular expression: \texttt{[a-zA-Z\_\$][a-zA-Z\textbackslash d\_\$]*(?:\textbackslash .{}[a-zA-Z\_\$][a-zA-Z\textbackslash d\_\$]*)+}}.
The strategies evaluated are: \texttt{num} (numeric only), \texttt{num \& path}, \texttt{num \& classpath}, and all (with all three types combined).
As demonstrated in Figure~\ref{fig:rq3-2}, the \texttt{num \& path} strategy is the optimal configuration for \texttt{LogFold}. 
This is primarily evidenced by the average CR peaking at 89.519 under this setting. 
While using only numeric tokens provides a strong baseline, the inclusion of path tokens offers a consistent, incremental benefit across most systems.
Conversely, incorporating classpath tokens (in the \texttt{num \& classpath} and \texttt{all} strategies) introduces performance volatility for several datasets, leading to a less stable performance profile and a lower average CR. 
Therefore, we select \texttt{num \& path} as the default configuration for \texttt{LogFold}, as it delivers the best combination of high average performance and robust stability across diverse systems, making it ideal for real-world deployment.

\answerbox{
\textbf{Answer to RQ3:} 
\texttt{LogFold} is highly robust across a range of parameter selections and token strategies.
}

\begin{figure}[tbp]
    \centering
    \includegraphics[width=0.8\linewidth]{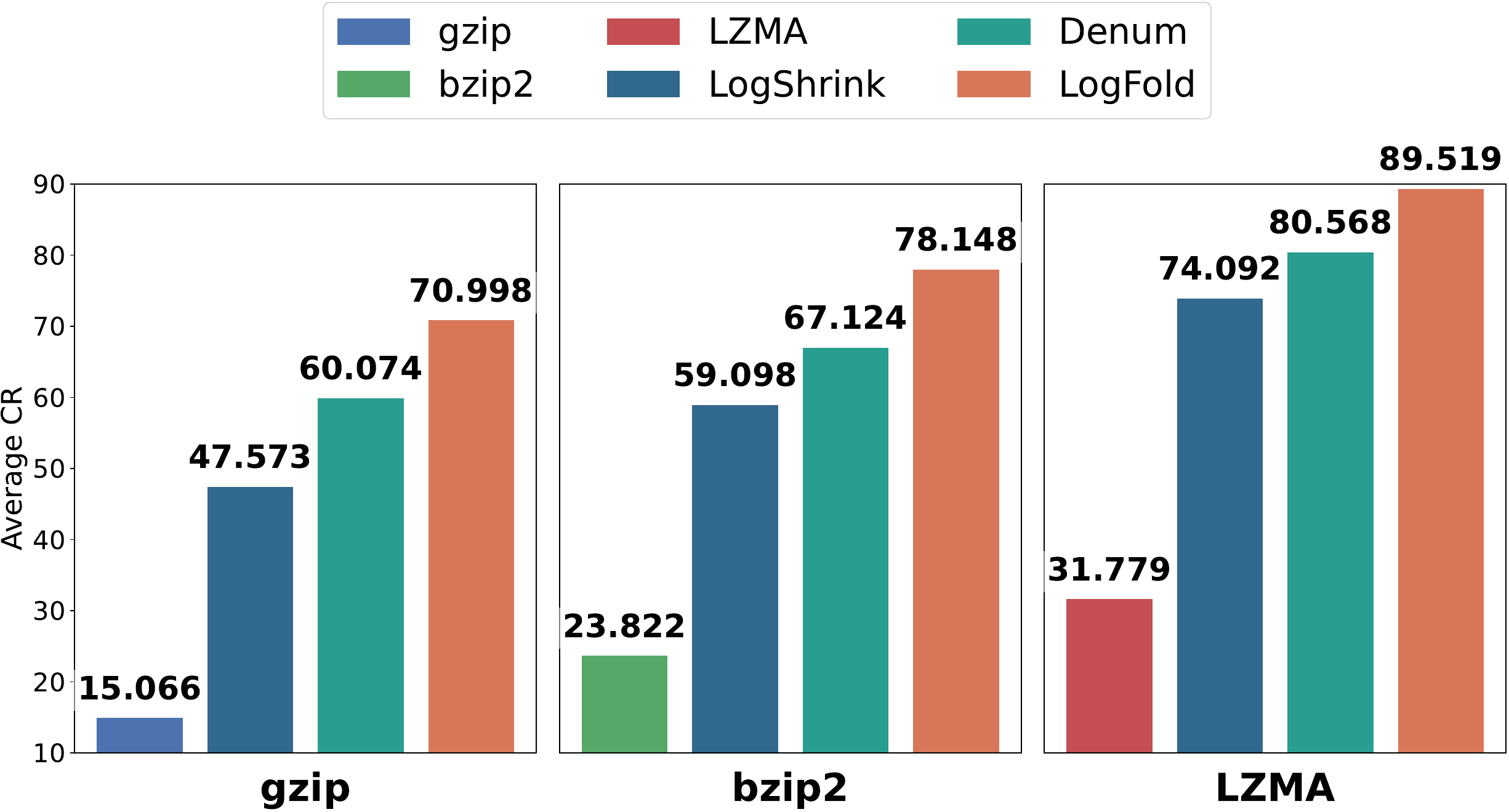}
    \vspace{-10pt}
    \caption{Average CR with different back-end general-purpose compressors.}
    \vspace{-10pt}
    \label{fig:rq3-3}
\end{figure}

\begin{figure}[tbp]
    \centering
    \includegraphics[width=0.7\linewidth]{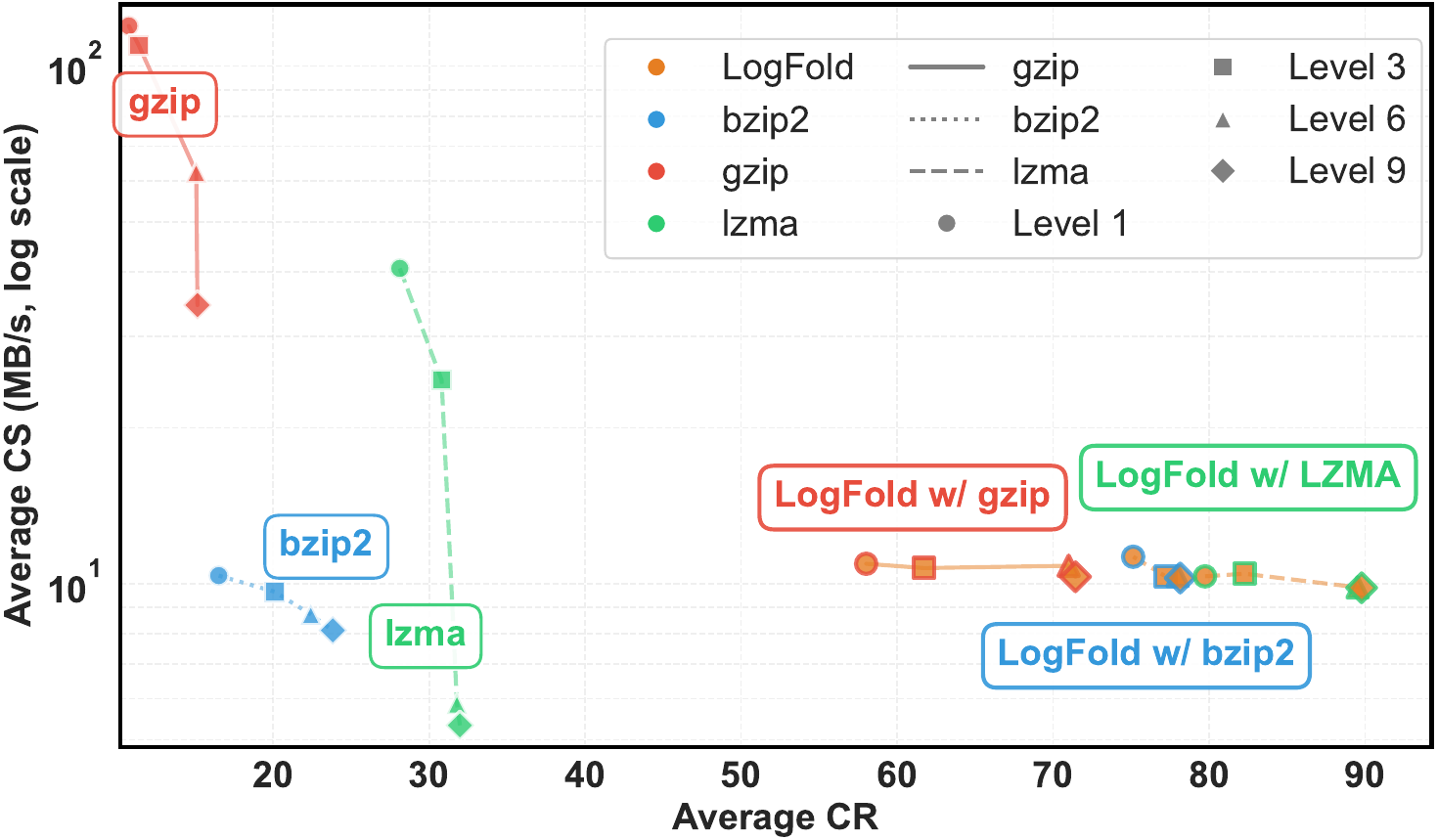}
    \vspace{-10pt}
    \caption{{Compression performance of different zip tools across various levels, highlighting the stability of \texttt{LogFold} which avoids the typical speed-versus-ratio trade-off.}}
    \label{fig:rq4-2}
\end{figure}
\subsection{RQ4: How generalizable is \texttt{LogFold} across different zip tools {with different compression levels}?}
As the final stage in typical log compression involves a general-purpose compressor, we investigate how different back-end compressors affect performance.
To this end, we compare \texttt{LogFold} with two leading log compressors, LogShrink and Denum, using three widely-adopted back-end algorithms: gzip, bzip2, and LZMA. 
The results depicted in Figure~\ref{fig:rq3-3} show that
\texttt{LogFold} consistently achieves the highest CR across multiple back-end compressors. It reaches 70.998 with gzip, 78.148 with bzip2, and a peak of 89.519 with LZMA, consistently outperforming the SOTA log compressors. 
This result confirms the generalizability of \texttt{LogFold} regardless of the final compression algorithm employed.

{Furthermore, we investigate the impact of different compression levels on the performance trade-off of \texttt{LogFold} between compression ratio and speed. Since standard compressors like gzip, bzip2, and LZMA offer multiple compression levels (typically 1-9), we select levels 1, 3, 6, and 9 as representative test points. The results of this analysis are depicted in Figure ~\ref{fig:rq4-2}.}
{As illustrated on the left side of the figure, the standalone tools exhibit conventional behavior: increasing the compression level (from Level 1 to Level 9) yields a progressively higher compression ratio but at the cost of a significant reduction in compression speed. This demonstrates the standard trade-off inherent in these algorithms. In contrast, when these tools are employed as backends for \texttt{LogFold}, this trade-off is virtually eliminated. The data points corresponding to levels 1, 3, 6, and 9 are tightly clustered. This indicates that the choice of the backend compression level has a negligible impact on both the final compression ratio and the overall processing speed of \texttt{LogFold}. It suggests that users can employ the highest compression level for the backend (e.g., Level 9) without sacrificing the compression gains achieved by \texttt{LogFold}.}
\answerbox{
\textbf{Answer to RQ4:} 
\texttt{LogFold} demonstrates strong generalizability,
consistently outperforming all baselines regardless of the back-end general-purpose compressor. {Its performance remains highly stable across different compression levels, effectively eliminating the usual speed-versus-ratio trade-off.
}}

\subsection{{RQ5: How does \texttt{LogFold} perform in log decompression?}}

\begin{table}[tbp]

\footnotesize
\centering
\caption{{Statistics of decompression speed of \texttt{LogFold} across 16 log datasets.}}
\vspace{-0.1in}
\begin{tabular}{cc|cc}
\toprule
\textbf{Dataset} & \textbf{DS (MB/s)} & \textbf{Dataset} & \textbf{DS (MB/s)} \\
\midrule
Android& 0.202 & Apache & 0.276 \\
BGL& 0.322 & Hadoop & 1.314 \\
HDFS & 0.132 & HealthApp & 0.299 \\
HPC & 0.119 & Linux & 0.450 \\
Mac& 0.212 & OpenSSH & 0.283 \\
OpenStack & 0.411 & Proxifier & 0.807 \\
Spark & 0.739 & Thunderbird & 0.194 \\
Windows & 1.418 & Zookeeper & 2.488 \\
\bottomrule
\end{tabular}
\label{tab:decompression_speed}
\end{table}

\begin{figure}[tbp]
    \centering
    \includegraphics[width=\linewidth]{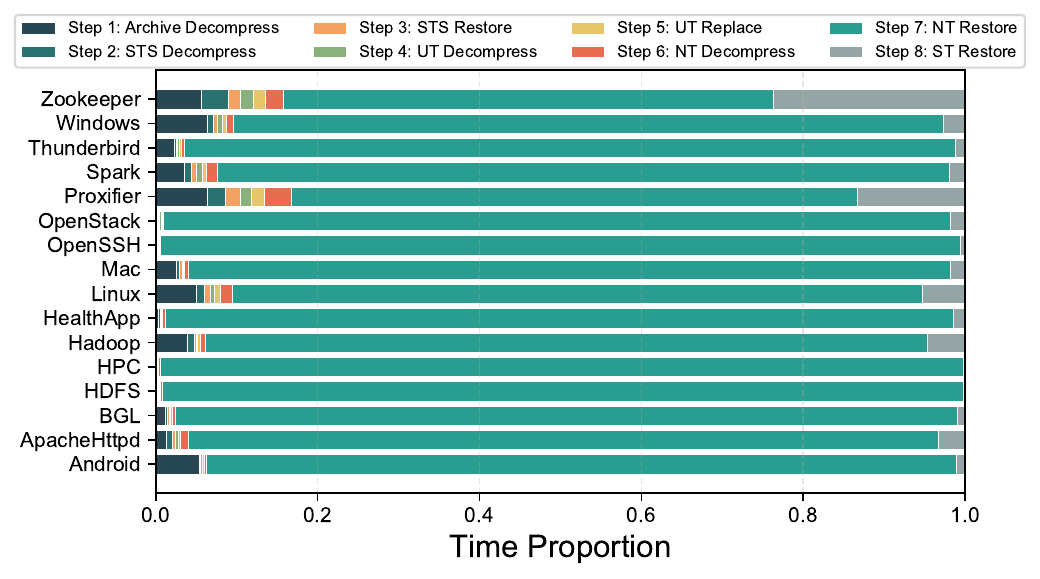}
    \vspace{-0.25in}
    \caption{{Time proportion of different steps in decompression.}}
    \label{fig:rq5}
    \vspace{-0.2in}
\end{figure}

{We demonstrate the decompression performance of \texttt{LogFold}, Table~\ref{tab:decompression_speed} shows the result.}
{As illustrated, the decompression speed varies across datasets, ranging from 0.119 MB/s on HPC to 2.488 MB/s on Zookeeper.}
{To understand this performance, we profiled the decompression process and measured the time spent on each step in Figure~\ref{fig:rq5}. Our profiling reveals that the primary bottleneck is not the backend decompression (e.g., LZMA) but specifically the restoration of numeric tokens (Step 7). This single step, which reconstructs the original numeric values, consumes the majority of the execution time. This computationally intensive process is a necessary trade-off for guaranteeing lossless restoration and achieving \texttt{LogFold}'s high compression ratios. }

\answerbox{
\textbf{{Answer to RQ5:}} 
{\texttt{LogFold}'s decompression speed is constrained by the computational overhead of its numeric token restoration (Step 7). This bottleneck is a necessary trade-off to ensure both lossless data recovery and superior compression ratios.}
}

\section{{Discussion}} 
\subsection{{Practical Utility}}
{The practical utility of \texttt{LogFold} is fundamentally linked to the inherent trade-off between compression ratio and speed~\cite{yao2020study,guerra2025learned,yu2024unlocking}. \texttt{LogFold} realizes its potential in scenarios that prioritize storage efficiency. These scenarios encompass long-term archiving, where infrequent access makes minimizing storage cost paramount (e.g., S3 Glacier~\cite{s3}); centralized log aggregation, where high compression is necessary to reduce network bandwidth and the central repository's storage footprint~\cite{yao2020study,cooltier}. \texttt{LogFold} provides practical value in these storage-sensitive environments by simultaneously achieving a state-of-the-art compression ratio against nine baselines while maintaining the second-fastest speed among log-specific compressors.}

\vspace{-0.15in}
\subsection{{Threats to Validity}}
\textbf{Internal Threats.}
{\texttt{LogFold} programmatically identifies any non-alphanumeric character as delimiters. It is designed to enhance generalizability and reduce manual effort compared to using predefined sets~\cite{dai2020logram, li2024logshrink}. While highly adaptable, it can over-generalize. A lightweight pre-processing step could merge empirical knowledge to balance adaptability with automation.  For dynamic tokens, it currently employs predefined regular expressions.}
While effective in evaluations, it may not be suitable for all log formats
{and} may introduce potential bias based on engineers’ experience or expertise.
{One future mitigation strategy is to automate this process by exploring statistical modeling following sampling, thereby improving the adaptability of \texttt{LogFold} on unseen systems.}

\textbf{External Threats.}
The primary threat to external validity is the generalizability of our results. Although our evaluation spans a diverse benchmark of 16 public datasets, it may not encompass the full range of log characteristics in all real-world environments, especially proprietary large-scale industrial systems. Consequently, \texttt{LogFold}’s performance might vary when applied to logs outside the domains studied.

\vspace{-0.15in}
\section{Related Work}
\textbf{Log Management.}
System logs are invaluable artifacts, recording critical runtime information that provides deep visibility into system behavior. {Their value has driven the development of automated log analysis, which can be broadly categorized into upstream, midstream, and downstream tasks~\cite{he2021survey}. Upstream tasks focus on log statement generation~\cite{li2021deeplv,batoun2024literature,ding2023logentext}, while midstream tasks address log parsing~\cite{dai2023pilar,dai2020logram,he2017drain,huo2023semparser,li2023did,ma2024librelog} and its use in aiding compression~\cite{liu2019logzip,li2024logshrink,wei2021feasibility}. Specifically, Logram~\cite{dai2020logram} operates as a log parser by building dictionaries from frequent n-gram token sequences to extract templates, whereas \texttt{LogFold} is designed for compression, analyzing structured tokens to identify compressible patterns.}

{Downstream tasks encompass a wide range of critical applications, including software testing~\cite{chen2018automated,chen2023exploring,xu2024mitigating}, system troubleshooting~\cite{tian2025ssdalog,he2025weakly,mariani2008automated,vervaet2021monilog}, and anomaly detection~\cite{shan2024face,huo2023autolog,liu2024logprompt,xu2009detecting,yang2024try}. }
In the current era, this field is being further advanced by Large Language Models (LLMs), which enable a new generation of sophisticated log analysis tools~\cite{zhang2024metalog,zhang2024end,shan2024face,he2024llmelog,jiang2025l4}. A common requirement for both traditional and modern LLM-based approaches is their reliance on vast quantities of log data to uncover meaningful insights. Consequently, the ability to efficiently store and manage these large volumes of logs has become a critical prerequisite for effective and scalable log analysis.

\textbf{Log Compression.}
A variety of specialized log compressors have been proposed for effective log storage~\cite{christensen2013adaptive,lin2015cowic,feng2016mlc,yao2021improving,rodrigues2021clp,wang2024muslope,yao2020study}.
Many of these methods leverage the inherent structure of logs, which consist of static templates and dynamic parameters. For example, LogZip\cite{liu2019logzip} employs iterative clustering to separate logs into templates and parameters for individual compression. Building on this paradigm, LogReducer\cite{wei2021feasibility} and LogShrink\cite{li2024logshrink} refine this by applying advanced optimizations to numerical data and latent structural patterns, respectively. A related strategy involves columnar processing, as seen in Cowic\cite{lin2015cowic} and LogBlock~\cite{yao2021improving}, which splits messages into fields and compresses each column independently.
{While \texttt{LogFold} also employs columnar processing, it differs from {LogBlock} by transposing sub-token matrices from tokens sharing a delimiter skeleton, rather than transposing entire columns from a parsed log matrix. This design enables \texttt{LogFold} to operate at a more granular sub-token level, capturing patterns among structurally similar tokens that may span different log templates and original column positions.} 
Other techniques bypass explicit template parsing.
LogArchive~\cite{christensen2013adaptive} and MLC~\cite{feng2016mlc} group similar log entries into buckets using similarity functions before compression. Denum~\cite{yu2024unlocking} offers another non-parsing alternative by first extracting and compressing all numeric tokens from the raw stream. 
ELISE~\cite{ding2021elise} represents a neural approach, converting all log content into a numerical representation for compression with a deep neural network.
CLP~\cite{rodrigues2021clp} and LogGrep~\cite{wei2023loggrep} efficiently query the compressed log data.

\section{Conclusion}
In this paper, we address the limitations of existing log compressors, which often fail to fully exploit redundancies within structured tokens and lack a fine-grained, type-aware encoding strategy for diverse tokens. We introduce \texttt{LogFold}, a novel, structured token-aware compressor that processes different token types with specialized components to achieve superior performance. Evaluations on 16 public datasets demonstrate that \texttt{LogFold} achieves a state-of-the-art average compression ratio of 89.519, outperforming existing baselines by 11.11\% at a practical speed. Our ablation study confirmed that this effectiveness is driven by its synergistic design, while sensitivity analyses verified its robustness. \texttt{LogFold} thus represents a significant advancement in log compression, offering a powerful and reliable solution for enhancing log compression.

\begin{acks}
We sincerely thank the anonymous reviewers for their critical and constructive evaluations.
This work was supported by the National Key Research and Development Program of China (2023YFB2704100) and the Singapore Ministry of Education (MOE) Academic Research Fund (AcRF) Tier 1 grant (Project ID: 24-SIS-SMU-082). Any opinions, findings and conclusions or recommendations expressed in this material are those of the author(s) and do not reflect the views of the Ministry of Education, Singapore.
\end{acks}

\newpage
\balance
\bibliographystyle{ACM-Reference-Format}
\bibliography{main}

\end{document}